\documentclass[a4paper]{jpconf}
\usepackage{graphicx}
\begin{document}
\title{First-principles band structure and FLEX approach to the pressure effect on $T_c$ of the cuprate superconductors}

\author{Hirofumi Sakakibara$^{1}$, Katsuhiro Suzuki$^{1}$, Hidetomo Usui$^{1}$, Kazuhiko Kuroki$^{1}$, Ryotaro Arita$^{2,5}$, Douglas J. Scalapino$^{3}$ and Hideo Aoki$^{4}$}

\address{$\rm ^1$Department of Engineering Science, The University of Electro-Communications, Chofu, Tokyo 182-8585, Japan}
\address{$\rm ^2$Department of Applied Physics, The University of Tokyo, Hongo, Tokyo 113-8656, Japan}
\address{$\rm ^3$Physics Department, University of California, Santa Barbara, California 93106-9530, USA}
\address{$\rm ^4$Department of Physics, The University of Tokyo, Hongo, Tokyo 113-0033, Japan}
\address{$\rm ^5$JST, PRESTO, Kawaguchi, Saitama 332-0012, Japan}

\ead{hiro$\_$rebirth@vivace.e-one.uec.ac.jp}

\begin{abstract}

High-temperature cuprate superconductors have been
 known to exhibit significant pressure effects. 
 In order to fathom the origin of why and how $T_c$ is affected
 by pressure, we have recently studied the pressure effects
 on $T_c$ adopting a model that contains two copper $d$-orbitals 
derived from first-principles
 band calculations, where the $d_{z^2}$ orbital is 
considered on top of the usually considered $d_{x^2-y^2}$
 orbital. In that paper, we have identified two origins for the $T_c$ 
enhancement under hydrostatic pressure:
 (i) while at ambient pressure the smaller 
the hybridization of other orbital
 components the higher the $T_c$, an application of pressure acts 
to reduce the multiorbital mixing on the
 Fermi surface, which we call the orbital distillation effect, and 
(ii) the increase of the band width with
 pressure also contributes to the enhancement. 
In the present paper, we further elaborate the two points.
 As for point (i), while 
the reduction of the apical oxygen height under pressure 
tends to increase the $d_{z^2}$  mixture, hence to lower $T_c$, 
here we show that this effect is strongly reduced in {\it bi-layer} materials
 due to the pyramidal coordination of oxygen atoms. 
As for point (ii), we show that the enhancement of
 $T_c$ due to the increase in the band width is caused by the 
effect that the many-body renormalization arising from the self-energy is reduced. 

\end{abstract}

\section{Introduction}

Although many kinds of superconductors have been discovered, 
the superconducting transition temperature $T_c$ 
of the cuprate superconductors  still remains to be 
the highest, and the possibility of further enhancing $T_c$  
still attracts much attention. 
To achieve higher $T_c$, it is important to understand the 
key parameters that control $T_c$, and 
from this viewpoint, 
there have been studies that have focused on the correlation 
between $T_c$ and the lattice parameters such 
as the in-plain bond length($l$)\cite{Jorgensen,Bianconi}
or the apical oxygen 
height($h_{\rm O}$)
\cite{Maekawa,Andersen,Feiner,Pavarini,Kotliar,Weber,Takimoto,PRL,PRB}  
measured from the CuO$_2$ planes.
Theoretically, we have studied the material dependence of $T_c$ 
in refs.\cite{PRL,PRB}, and introduced a two-orbital model that 
takes into account the $d_{x^2-y^2}$ and $d_{z^2}$ Wannier orbitals.
In most of the theories of the cuprates, 
only the $d_{x^2-y^2}$ (and the hybridized oxygen $p$) orbital 
is considered, but actually 
it has been noticed from the early days that in 
(La,Sr)$_2$CuO$_4$, which has small $h_{\rm O}$ and relatively low $T_c$, 
there is a strong mixture of the $d_{z^2}$ orbital component 
near the Fermi level\cite{Shiraishi,Eto,Freeman}.
In fact, nowadays there have been more studies that focus on the 
$d_{z^2}$ orbital or the related apical oxygen
\cite{Kotliar,Weber,Millis,Fulde,Honerkamp,Mori}.
In refs.\cite{PRL,PRB}, we have shown that 
this mixture of the $d_{z^2}$ orbital component near the Fermi level 
works destructively 
against $d-$wave superconductivity, and hence 
this is the main reason of the material dependence of $T_c$.

Studying the pressure effect on $T_c$ is an {\it in situ} way of 
attacking the problem of the correlation between $T_c$ and the lattice 
structure. It is well known that in most of the cuprates, application of 
pressure enhances $T_c$\cite{Klehe,Gao}. 
On the other hand, 
it has been revealed that the pressure effect exhibits strong 
anisotropy\cite{Hardy,Gugenberger,Meingast}.
In our recent theoretical study on the  pressure effect in the 
cuprates, it has been revealed that in addition to the $d_{z^2}$ effect, 
the roundness of the Fermi surface,
which is controlled by Cu-$4s$-$d_{x^2-y^2}$ hybridization,  
and also the band width are important parameters that 
governs $T_c$ under pressure\cite{pressure}.

 In the present paper, after briefly reviewing the main results 
obtained in ref.\cite{pressure}, 
we study more closely the effect of the band width 
by applying the fluctuation exchange 
approximation(FLEX)\cite{Bickers,Dahm,Kontani} to the two-orbital model.
Secondly, we will explain the difference of the $d_{z^2}$ orbital effect 
between the multi-layer and single-layer systems under hydrostatic pressure.

\section{Calculation Method}

\subsection{Determination of the Crystal Structure Under Pressure}

We first obtain the crystal structure of the single-layer cuprates 
La$_2$CuO$_4$ and Hg$_2$BaCuO$_4$ under ambient pressure.
Namely, we calculate the total energy by 
first principles calculation\cite{wien2k} 
varying the lattice constants, and fit the result by the 
standard Burch-Marnaghan formula\cite{Birch} 
to obtain the structure at the most stable point.
From such calculation, we can obtain the crystal structure within 
one percent discrepancy from the lattice constants determined 
experimentally\cite{La-st,Hg-st}.
To obtain the lattice structure under hydrostatic pressure, 
we optimize the Poisson's ratio 
and atomic position by first principles
reducing the cell volume to 95$\%$ or 90$\%$ of the lattice structure 
under ambient pressure.
Since the compressibility in the cuprates is known to be 
about $\sim 0.01$ GPa$^{-1}$\cite{Balagurov},
$V=0.9V_0$ corresponds to a pressure of about 10 GPa.

\subsection{Construction of the two-orbital model and FLEX approximation}

Using the obtained crystal structure under ambient or hydrostatic pressure, 
we construct maximally localized Wannier orbitals\cite{w2w,MaxLoc} to 
extract the hopping parameters 
of the $d_{x^2-y^2}$-$d_{z^2}$ two-orbital model\cite{PRL}.
As for the electron-electron interactions, 
we consider the on-site intraorbital Coulomb repulsion $U$, 
interorbital repulsion $U'$, the Hund's coupling
 $J$ and the pair-hopping $J'$. 
Here we also keep the orbital SU$(2)$ requirement, $U-U'=2J$.
We set $U=3.0$ eV, $U'=2.4$ eV and $J=J'=0.3$ eV unless mentioned 
otherwise.
Estimates of $U$ for the cuprates is 
$7-10t$, $t\simeq 0.45$eV
(namely, $U$ is about $3\simeq 4.5$ eV), and $J(J')\simeq 0.1U$,
so the values adopted here are within the widely accepted range. 
  
We apply FLEX to this model to 
obtain the Green's function renormalized by the 
many-body self-energy correction.
In FLEX, we define the spin and charge susceptibilities as follows;

\begin{equation}
\hat{\chi}_s(q)=\frac{\hat{\chi}^0(q)}{1-\hat{S}\hat{\chi}^0(q)} ,
\end{equation}
\begin{equation}
\hat{\chi}_c(q)=\frac{\hat{\chi}^0(q)}{1+\hat{C}\hat{\chi}^0(q)} ,
\end{equation}
where $q \equiv (\vec{q},i\omega_n)$, 
the irreducible susceptibility is 
\begin{equation} 
\chi^0_{l_1,l_2,l_3,l_4}(q) =\sum_q G_{l_1l_3}(k+q)G_{l_4l_2}(k)
\end{equation}
with the interaction matrices 

\begin{equation}
S_{l_1l_2,l_3l_4}
=\left\{\begin{array}{cc}
U, &\;\; l_1=l_2=l_3=l_4\\ 
U',&\;\; l_1=l_3\neq l_2=l_4\\
J,&\;\; l_1=l_2\neq l_3=l_4\\
J',&\;\; l_1=l_4\neq l_2=l_3 ,
\end{array}  \right.
\end{equation}

\begin{equation}
C_{l_1l_2,l_3l_4}
=\left\{\begin{array}{cc}
 U &\;\; l_1=l_2=l_3=l_4\\ 
-U'+J & \;\; l_1=l_3\neq l_2=l_4\\
2U'-J,&\;\; l_1=l_2\neq l_3=l_4\\
J'& \;\; l_1=l_4\neq l_2=l_3 ,
\end{array}  \right.
\end{equation}
here, $l_1,l_2$ are orbital indices.
Considering the self-energy correction originating 
from these susceptibilities,
we solve the Dyson equation in a self-consistent manner.
Then the renormalized Green's function is substituted 
to the linearized Eliashberg equation for superconductivity,

\begin{eqnarray}
\lambda \Delta_{ll'}(k) = -\frac{T}{N}\sum_q
\sum_{l_1l_2l_3l_4}V_{l l_1 l_2 l'}(q) 
G_{l_1l_3}(k-q)\Delta_{l_3l_4}(k-q)
G_{l_2l_4}(q-k).
\end{eqnarray}

The maximum eigenvalue $\lambda$ in the above equation reaches unity at 
the superconducting transition temperature $T=T_c$, 
so that $\lambda$ calculated at a fixed temperature 
can be used as a measure for $T_c$.
Here, we calculate $\lambda$ at $T=0.01$ eV for La and  $T=0.03$ eV for Hg. 
The reason for this is because $T_c\sim 100$ K of Hg is about 
three times larger than for La with $T_c\sim 40$ K\cite{Eisaki}.

\section{Results and Discussions: effect of band width}

\subsection{Orbital Distillation}

\begin{figure}[h!]
\includegraphics[width=12cm]{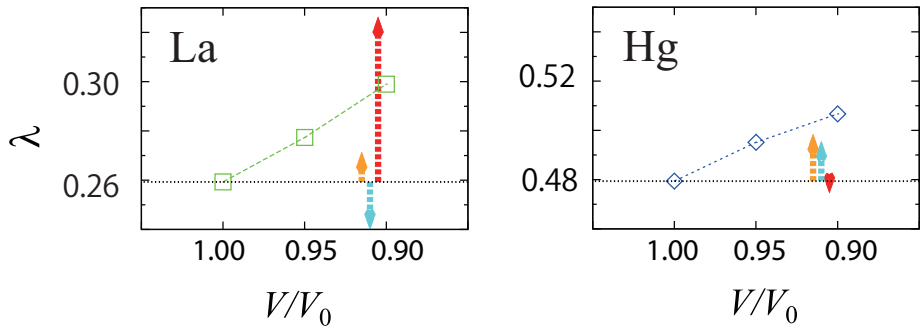}
\caption{Eigenvalue $\lambda$ of the Eliashberg equation plotted against 
the volume compression $V/V_0$.
Arrows indicate the contribution to the $\lambda$ variance from the 
parameters $\Delta E$, $r_{x^2-y^2}$ and $W$ (see text). 
}
\label{hydro}
\end{figure}

In Fig.\ref{hydro}, $\lambda$ is depicted as a function of the unit 
cell volume.
$\lambda$ increases in La and in Hg cuprates with hydrostatic 
pressure, and this agrees with the well-known experimental 
result that $T_c$ goes up 
monotonically under pressure in the cuprates\cite{Klehe,Gao}.
To understand this result, we have introduced 
three important parameters, the level offset $\Delta E$,
the roundness of Fermi surface ($r_{x^2-y^2}$ defined below) and the band 
width $W$\cite{pressure}.
To decompose the pressure effect into the contribution from the 
variation of these parameters,
we vary each parameter ``by hand'' from the original value at $V=V_0$ to 
the value at $V=0.9V_0$ separately, and calculate the variation in $\lambda$.
This result is also depicted in Fig.\ref{hydro} as the length of arrows.
$\Delta E$ is the on-site energy difference between  $d_{x^2-y^2}$ and
$d_{z^2}$ Wannier orbitals in the two-orbital model, so this is a measure of
the $d_{z^2}$ orbital effect, which has been found to work destructively 
against $d$-wave superconductivity in our previous study\cite{PRL}.
The effect of $\Delta E$ is dominant in La compound because La
 originally has small $\Delta E$ thereby suppressing $T_c$, 
so its increase under pressure is effective for the $T_c$ enhancement.
Note that in La system, $\Delta E$ increases under pressure 
because the absolute magnitude 
of the crystal field increases in spite of the reduction of $h_{\rm O}/l$.
On the other hand, in the Hg compound the pressure effect through $\Delta E$
is negligible because $\Delta E$ is intrinsically large.
Instead, the other two parameters are effective.
$r_{x^2-y^2}$ is defined as $r_{x^2-y^2}\equiv (|t_2|+|t_3|)/|t_1|$, 
where $t_i$ is the $i$-th neighbor hopping within the $d_{x^2-y^2}$ orbital.
The roundness of Fermi surface is enhanced by the increase of this value, 
and $t_2$ and $t_3$ are mostly mediated 
by the Cu-$4s$ orbital (which is effectively included in the two Wannier 
orbitals in the present model) with the path 
of $d_{x^2-y^2} \rightarrow 4s \rightarrow d_{x^2-y^2}$\cite{Andersen}.
It is known that the roundness of the Fermi surface works 
against the spin-fluctuation-mediated 
superconductivity\cite{Scalapino}, 
so the reduction of the $4s$ effect enhances $T_c$.
The $4s$ orbital effect is reduced by hydrostatic pressure 
because the energy level offset between the 
$d_{x^2-y^2}$ and the $4s$ orbital 
is enhanced when the oxygen ligands approach Cu.
The effect of $d_{z^2}$ and $4s$ orbitals put together, 
we can say that $T_c$ increases when the 
main band has more pure $d_{x^2-y^2}$ component,
namely, when the ``orbital distillation'' takes place.

\subsection{The Effect of Band Width $W$}

\begin{figure}[h]
\includegraphics[width=14cm]{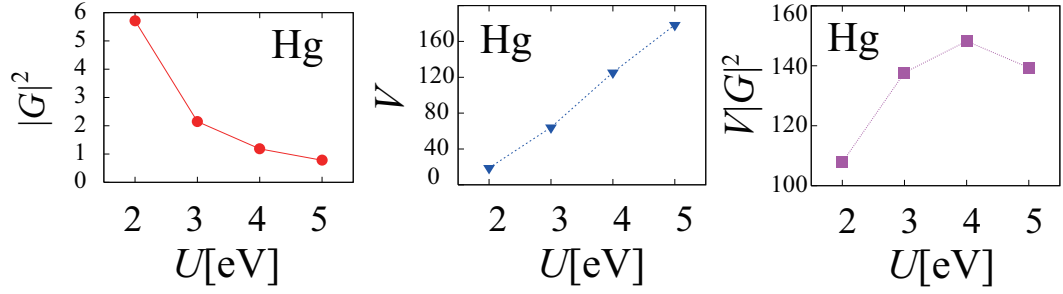}
\caption{$U$ dependence of the FLEX calculation result for the Hg compound. 
(left) the absolute value of the renormalized
 Green's function squared $|G|^2$ at $(\vec{k},i\omega)=(\pi,0,i\pi k_BT)$,
(middle) the effective pairing interaction 
$V$ at $(\vec{q},i\omega)=(\pi,\pi,0)$ 
and (right) the product $V|G|^2$. 
}
\label{vgg}
\end{figure}

Now, we turn to the first main topic of this paper, i.e., 
the effect of the band width $W$.
The band width $W$ is defined as the energy range 
between the top and the bottom of the main band.
It is evident that the band width is controlled 
by the in-plain bond length $l$,
so that hydrostatic pressure enhances $W$.
In Fig.\ref{hydro}, we can see that the increase in $W$ 
results in an enhancement of $\lambda$ in Hg, while the 
opposite occurs for La.
The reason for this can be understood as follows.
Let us first start with the $U$ dependence of $T_c$ for a fixed band width.
In the top of Fig.\ref{vgg}, we show the absolute value of the renormalized
 Green's function squared $|G|^2$ in the Hg compound 
at $(\vec{k},i\omega)=(\pi,0,i\pi k_BT)$ 
for several values of $2<U<5$ eV.
The value of $|G|^2$  monotonically decreases 
with larger $U$ because the self-energy, 
which increases with $U$,  suppresses $|G|^2$.
On the other hand, 
the pairing interaction shown in the middle increases with $U$ 
because the spin fluctuations develop monotonically.
Consequently, $V|G|^2$ shown on the right exhibits a maximum at 
a certain $U$. Since $V|G|^2$ can be 
considered as a rough measure of the eigenvalue of the Eliashberg equation 
for $d-$wave superconductivity, $\lambda$ (and thus $T_c$) 
is expected to be {\it maximized} around a certain $U_{\rm max}$.
%This picture is totally different from BCS limit\cite{Allen}, particular in 
%intermediately or strongly correlated system, such as cuprates.
%Based on such point of view, 
%applied pressure may enhance $\lambda$(and hence $T_c$)
%if the interaction $U$ has certain large value.

\begin{figure}[h]
\includegraphics[width=12cm]{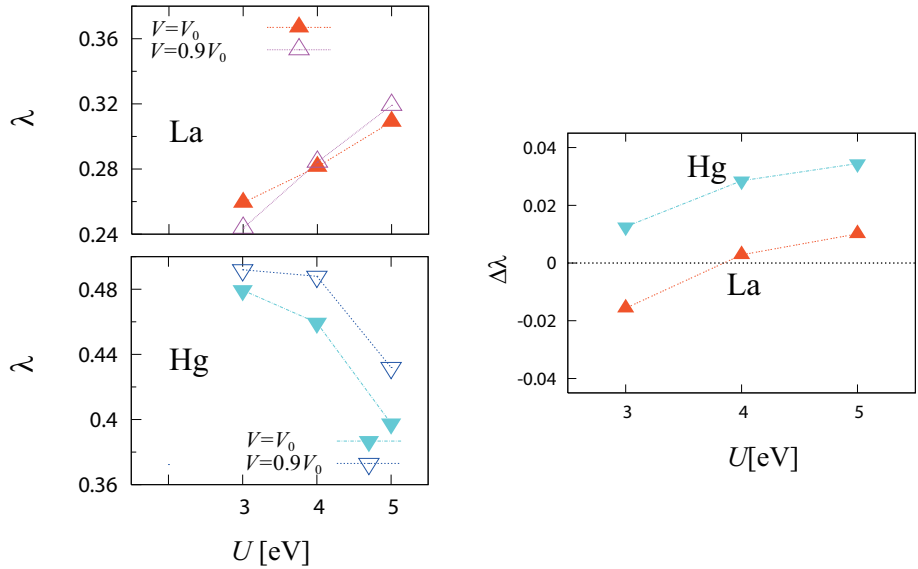}
\caption{(Left) $\lambda$ against $U$ for La (top) and Hg (bottom).
Filled (opened) symbols displays the result for $V=V_0$($V=0.9V_0$).
(Right) $\Delta\lambda$ (increment in 
$\lambda$ when $W$ is increased up to its value at $V=0.9V_0$) against 
$U$.
}
\label{dL}
\end{figure}

In the left panels of Fig.\ref{dL}, we show the $U$ dependence of 
$\lambda$  for $3<U<5$ eV 
for the two compounds with $V=V_0$ and $0.9V_0$.
We can see that $U<5$ eV lies on the left side of $U_{\rm max}$ for 
La, while 3 eV$<U$ is on the right side of $U_{\rm max}$ in Hg.
This means that larger values of $U$ is necessary for La to be in the 
``strongly correlated regime''.
This is because electrons can avoid the strong intraorbital 
repulsion within $d_{x^2-y^2}$ orbitals 
by using the $d_{z^2}$ orbital degrees of freedom in 
materials with small $\Delta E$.

In the right panel of Fig.\ref{dL}, we show the $U$ dependence of 
the increment $\Delta\lambda$ of the eigenvalue $\lambda$
induced by the increase of  $W$ under pressure. 
In this calculation, 
$W$ is increased  ``by hand'' up to its value at $V=0.9V_0$, 
while the other two parameters are fixed at their original values.
From this figure, we can see that the increase in $W$ always enhances 
$T_c$ in the Hg cuprate within the realistic $U$ range.
For the La cuprate on the other hand, $\Delta\lambda$ is negative 
for small values of $U$, and this is the  reason why 
$W$ affects superconductivity in opposite ways between La and Hg in Fig.\ref{hydro}.
For larger values of $U$, however, $\Delta\lambda$ becomes positive 
even for La.

\begin{figure}[h]
\includegraphics[width=12cm]{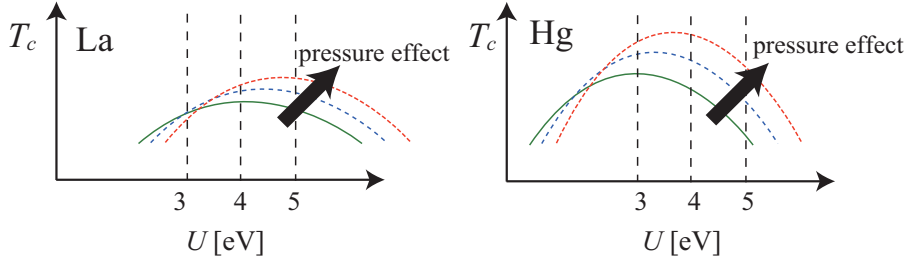}
\caption{
Schematic figure of $T_c$ variation induced by applied pressure.
Each line is a $U$ vs. $T_c$ plot for fixed $W$, and $W$ is larger for 
solid green $\rightarrow$ dashed blue $\rightarrow$ dotted red.
}
\label{scheme}
\end{figure}

To see the effect of the band width more clearly, 
we provide in Fig.\ref{scheme} a schematic 
band width dependence of the $U$ vs $T_c$ plot.
Namely, $T_c$ should depend on $U$ and $W$ essentially in the form $W f(U/W)$, 
where $f$ is a certain function that gives the overall 
dependence of $T_c$ against the electron correlation strength.
Therefore, as the pressure is applied, $W$ increases so that 
$U_{\rm max}$ (peak position of the curve)
also increases accordingly keeping $U_{\rm max}/W$ constant.
At the same time, the absolute value of the maximized $T_c$ is 
enhanced by the application of the pressure because the entire energy scale 
increases (i.e., both $U_{\rm max}$ and $W$ are enhanced).

\section{The $d_{z^2}$ Orbital Effect: Multi-layer vs Single-layer Systems}

So far, we have concentrated on single-layer systems. 
In this section, we will discuss the difference of the pressure 
effect through $\Delta E$ between single and multi-layer systems.
Here, we focus on the comparison between single- and bi-layer 
La, Hg, Tl, Bi, and Y cuprates.
We consider $\Delta E_d$, the energy difference  
between the $d_{x^2-y^2}$ and the $d_{z^2}$ orbital 
in the $d$-$p$ model which considers all of the Cu $3d$ and 
O $2p$ orbitals explicitly.
In the $d$-$p$ model, the basis functions 
for the hopping part are close to 
the atomic orbitals, so $\Delta E_d$ can be considered 
as the energy difference between the 
$d_{x^2-y^2}$ and the $d_{z^2}$ atomic orbitals.
$\Delta E$ is the energy difference between the 
$d_{x^2-y^2}$ and the $d_{z^2}$ Wannier orbitals which 
effectively take the O$2p$  orbitals into account, 
so $\Delta E_d$ and $\Delta E$ are positively correlated\cite{PRB}.

\begin{figure}[h]
\includegraphics[width=8cm]{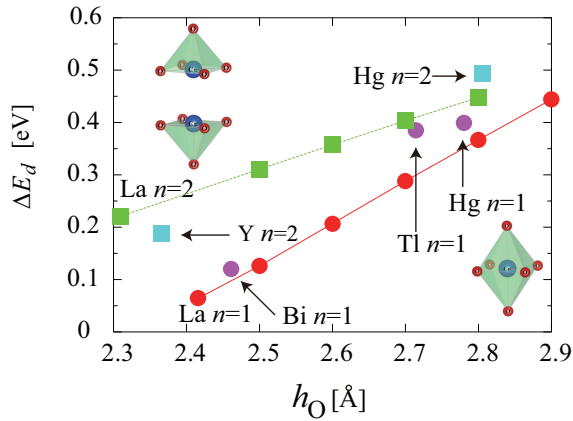}
\caption{Relationship between the apical oxygen height $h_{\rm O}$[\AA] and
the level difference $\Delta E_d$ [eV]. The squares(circles) show the value of $\Delta E_d$ 
in each bi-layer(single-layer) system(the materials are indicated by arrows).
The dashed(solid) line presents the calculation result when $h_{\rm O}$ is varied hypothetically
in bi-layer (single-layer) La system by hand.
}
\label{ho-ed}
\end{figure}

In Fig.\ref{ho-ed}, we show the relationship between 
the apical oxygen height $h_{\rm O}$ and $\Delta E_d$.
We can see that, while 
$\Delta E_d$ is positively correlated with the apical oxygen height
in both single- and bi-layer systems as expected, 
$\Delta E_d$ is overall significantly greater 
in the bi-layer systems than 
in the single-layer systems.  
This is because bi-layer cuprates take pyramidal coordination 
of the oxygen ligands,  while the single-layer cuprates take an 
octahedral one. 
Hence the effect of the apical oxygen should be weaker in the 
bi-layer systems, so that the ``effective $h_{\rm O}$'' is larger.
As we have seen  in the previous section, $\Delta E$ (and hence the $d_{z^2}$ orbital) 
plays minor role in the $T_c$ enhancement under pressure in 
materials with large $\Delta E$, so the effect of 
 $\Delta E$ should become  less relevant 
as the number of the layers increases.
This can be considered as one reason why $T_c$ increases in spite of 
somewhat larger reduction in $h_{\rm O}$ 
in bi-layer system than in single-layer system 
under hydrostatic pressure\cite{Jorgensen}.

\section{Conclusion}

In conclusion, we have performed 
first-principle calculation+FLEX study to analyze 
the pressure effect on $T_c$ of the cuprates.
To explain the $T_c$ enhancement induced by hydrostatic pressure,
we introduce three important parameters, level offset $\Delta E$, 
roundness of the Fermi surface $r_{x^2-y^2}$ 
and the band width $W$.
$\Delta E$ and $r_{x^2-y^2}$ are the measure of "orbital purity" 
in the main band, and $T_c$ is 
enhanced by ``orbital distillation''.
We have shown that hydrostatic pressure enhances the distillation, thereby 
enhancing $T_c$. 
We have also analyzed the effect of the band width $W$ in detail. 
It has been shown that $\lambda$ (and hence $T_c$) 
is maximized around a certain $U$, which depends on the band width as 
well as $\Delta E$. For materials with large $\Delta E$ such as the 
Hg compound, the increase of $W$ under pressure leads to an enhancement of 
$\lambda$ for realistic values of $U$.

We also discuss the difference between bi- or single-layer systems 
from the view point of $\Delta E$.
Even when $h_{\rm O}$ is suppressed under pressure 
in bi-layer systems, the effect on  $T_c$ should be small 
because of the pyramidal coordination of the oxygen ligands.

\section{Acknowledgments}

The numerical calculations were performed at the Supercomputer Center, 
ISSP, University of Tokyo. This study has been supported by 
Grants-in-Aid for Scientific Research from JSPS
(Grants No. 23340095, RA;
No. 23009446, HS; No. 21008306, HU; and
No. 22340093, KK and HA). RA acknowledges financial support from 
JST-PRESTO.
DJS acknowledges support from the Center for 
Nanophase Material Science at Oak Ridge National Laboratory.


\section*{References}
\begin{thebibliography}{99}
\bibitem{Jorgensen} %4
J.D. Jorgensen, D.G. Hinks, O.Chmaissem, D.N. Argyiou, J.F. Mitchell, and B. Dabrowski, in 
{\it Lecture Notes in Physics}, {\bf 475}, p.1 (1996);
J.D. Jorgensen, D.G. Hinks, O.Chmaissem, D.N. Argyiou, J.F. Mitchell, and B. Dabrowski,
in {\it Recent Developments in High Temperature Superconductivity}, edited by 
J.Klamut, B.W. Veal, B.M. Dabrowski, P.W. Klamut and M.Kazimierski(Springer, Poland, 1995), pp.1-15
\bibitem{Bianconi} % 5
A. Bianconi, G. Bianconi, S. Caprara, D. Di Castro, 
H. Oyanagi and N. L. Saini, 
J. Phys.: Condens. Matter {\bf 12}, 10655 (2000); 
A. Bianconi, S. Agrestini, G. Bianconi, D. Di Castro, and 
N. L. Saini, J. Alloys Compd. {\bf 317-318}, 537 (2001); 
N. Poccia, A. Ricci and A. Bianconi, Adv. Condens. 
Matter Phys. {\bf 2010}, 261849 (2010). 
\bibitem{Maekawa} % 6
S. Maekawa, J. Inoue and T. Tohyama, in {\it The Physics 
and Chemistry of Oxide Superconductors}, edited by Y. Iye 
and H. Yasuoka (Springer, Berlin, 1992), Vol. {\bf 60}, 
pp. 105-115. 
\bibitem{Andersen} % 7
O.K. Andersen, A.I. Liechtenstein, O. Jepsen, and F. Paulsen,
 J. Phys. Chem. Solids {\bf 56}, 1573 (1995).
\bibitem{Feiner} %8
L.F. Feiner, J.H. Jefferson and R. Raimondi, Phys. Rev. Lett. {\bf 76}, 4939 (1996).
\bibitem{Pavarini} %9
E. Pavarini, I. Dasgupta, T. Saha-Dasgupta, O. Jepsen, and O. K. Andersen, Phys. Rev. Lett. {\bf 87}, 047003 (2001).
\bibitem{Kotliar} %10
C. Weber, K. Haule, and G. Kotliar, %K. Haule, and G. Kotliar, 
Phys. Rev. B {\bf 82}, 125107(2010).
\bibitem{Weber} %11
C. Weber, C. -H. Yee, K. Haule and G. Kotliar, 
arXiv:1108.3028. 
\bibitem{Takimoto} %12
T. Takimoto, T. Hotta and K. Ueda, Phys. Rev. B {\bf 69}, 104504 (2004).

\bibitem{PRL} H. Sakakibara, H. Usui, K. Kuroki, R. Arita, and H. Aoki, %2
Phys. Rev. Lett. {\bf 105}, 057003 (2010). 
\bibitem{PRB} %3
H. Sakakibara, H. Usui, K. Kuroki, R. Arita and H. Aoki, Phys. Rev. B {\bf 85}, 064501 (2012).
\bibitem{Shiraishi} %22    
K. Shiraishi, A. Oshiyama, N. Shima, T. Nakayama and H. Kamimura, Solid State Commun. {\bf 66},
629 (1988).
\bibitem{Eto} %23
H. Kamimura and M. Eto, J. Phys. Soc. Jpn. {\bf 59}, 3053 
(1990); M. Eto and H. Kamimura, J. Phys. Soc. Jpn. {\bf 60}, 2311 (1991).
\bibitem{Freeman} % 24 
A.J. Freeman and J. Yu, Physica B {\bf 150}, 50 (1988).
\bibitem{Millis} %16
X. Wang, H.T. Dang, and A. J.Millis, 
Phys. Rev. B {\bf 84}, 014530(2011).
\bibitem{Fulde} %17
L. Hozoi, L. Siurakshina, P. Fulde and J. van den Brink, 
Sci. Rep. {\bf 1}, 65 (2011). 
\bibitem{Honerkamp} %18 
S. Uebelacker and C. Honerkamp, Phys. Rev. B {\bf 85}, 
155122 (2012).
\bibitem{Mori}  %19
M. Mori, G. Khaliullin, T. Tohyama, and S. Maekawa, Phys. Rev. Lett. 101, 247003 (2008)

\bibitem{Klehe} %36
A.-K. Klehe, A. K. Gangopadhyay, J. Diederichs and J. S. Schilling Physica C {\bf 213} 266 (1993); {\bf 223} 121(1994).
\bibitem{Gao} %37
L. Gao, Y. Y. Xue, F. Chen, Q. Xiong, R. L. Meng, D. Ramirez, C. W. Chu, J.H Eggert, and H.K. Mao, 
Phys. Rev. B {\bf 50}, 4260 (1994).

\bibitem{Hardy} % 13
F. Hardy, N. J. Hillier, C. Meingast, D. Colson, Y. Li, N. Barisic, G. Yu, X. Zhao, M. Greven, and J. S. Schilling, Phys. Rev. Lett. {\bf 105}, 167002 (2010).
%(comment)In Y-123 system, the experimental result is out of the law
%;C. Meingast, {\it et al.}, Phys. Rev. Lett. {\bf 67}, 1634 (1991)
%;U. Welp, {\it et al.}, Phys. Rev. Lett. {\bf 69}, 2130 (1992).
\bibitem{Gugenberger} % 14 
F. Gugenberger, C. Meingast, G.Roth, K. Grube, V. Breit, T. Weber, H. Wuhl, S. Uchida, and Y. Nakamura, Phys. Rev. B {\bf 49}, 13137 (1994).
\bibitem{Meingast} % 15
C. Meingast, A. Junod and E. Walker, Physica C {\bf 272}, 106 (1996).

%\bibitem{Maier} %20
%Th. Maier, M. Jarrell, Th. Pruschke, and J. Keller, Phys. Rev. Lett. {\bf 85}, 1524 (2000).
%\bibitem{Kent} %21
%P. R. C. Kent, T. Saha-Dasgupta, O. Jepsen, O. K. Andersen, A. Macridin, T. A. Maier, M. Jarrell, and T. C. Schulthess, Phys. Rev. B {\bf 78}, 035132 (2008).


\bibitem{pressure}%1
H. Sakakibara, K. Suzuki, H. Usui, K. Kuroki, R. Arita, D.J. Scalapino and Hideo Aoki, 
Phys. Rev. B {\bf 86}, 134520 (2012)

\bibitem{Bickers} %25
N.E. Bickers, D.J. Scalapino, and S.R. White, 
Phys. Rev. Lett. {\bf 62}, 961 (1989).
\bibitem{Dahm} %26
 T. Dahm and L. Tewordt, Phys. Rev. Lett. {\bf 74}, 793 (1995).
\bibitem{Kontani}  % 27
K. Yada and H. Kontani, J. Phys. Soc. Jpn. {\bf 74}, 2161 (2005).

\bibitem{wien2k} %28
P. Blaha, K. Schwarz, G.K.H. Madsen, D. Kvasnicka, and J. Luitz, 
{\it Wien2k: An Augmented Plane Wave} + {\it Local Orbitals Program for Calculating Crystal Properties} (Vienna University of Technology, Wien, 2001).
\bibitem{Birch} %29
F. Birch, Phys. Rev. B {\bf 71}, 809(1947).

\bibitem{La-st} %30
J. D. Jorgensen, H. -B. Schuttler, D. G. Hinks, D. W. Capone, K. Zhang, and M. B. Brodsky and D. J. Scalapino, Phys. Rev. Lett. {\bf 58}, 1024 (1987).
%J. Yu {\it et al.},Phys. Rev. Lett. {\bf 58}, 1035 (1987).
\bibitem{Hg-st} %31
J.L. Wagner, P.G. Radaelli, D.G. Hinks, J.D. Jorgensen, J.F. Mitchell, B. Dabrowski, G.S. Knapp, 
M.A. Beno, Physica C {\bf 210}, 447 (1993).
%D. Downs, Phys. Rev. B {\bf 52}, 15627 (2009).
\bibitem{Balagurov} %32
A. M. Balagurov, D. V. Sheptyakov, V. L. Aksenov,
E. V. Antipov, S. N. Putilin, P. G. Radaelli and M. Marezio, Phys. Rev. B {\bf 59}, 7209 (1999). 
\bibitem{w2w} %33
J. Kunes, R. Arita, P. Wissgott, A. Toschi, H. Ikeda, and K. Held, Comp. Phys. Commun. {\bf 181}, 1888 (2010).
\bibitem{MaxLoc} %34
N. Marzari and D. Vanderbilt, Phys. Rev. B 
{\bf 56}, 12847 (1997); 
I. Souza, N. Marzari and D. Vanderbilt, 
Phys. Rev. B {\bf 65}, 035109 (2001).
The Wannier functions are generated by the code developed by
A. A. Mostofi, J. R. Yates, N. Marzari, I. Souza, and D. Vanderbilt,
http://www.wannier.org/.
\bibitem{Eisaki} %35
H. Eisaki, N. Kaneko, D. L. Feng, A. Damascelli, P. K. Mang, K. M. Shen, Z.-X. Shen, and M. Greven,
 Phys. Rev. B {\bf 69}, 064512(2004).

\bibitem{Scalapino} %38
For a review, see D.J. Scalapino in 
{\it Handbook of High Temperature Superconductivity}, Chapter 13, 
Eds. J.R. Schrieffer and J.S. Brooks (Springer, New York, 2007).
%\bibitem{Allen} %39
%P.B. Allen and P.C. Dynes, Phys. Rev. B {\bf 12}, 905 (1975).

%\bibitem{Takahashi}
%H. Takahashi, H. Shaked, B. A. Hunter, P. G. Radaelli, R. L. Hitterman, D. G. Hinks, and J. D. Jorgensen , Phys. Rev. B {\bf 50}, 3221 (1994). 
%\bibitem{Tanaka} K. Tanaka {\it et al.}, Phys. Rev. B {\bf 70}, 092503 (2004).
%\bibitem{Moriya} T. Moriya and K. Ueda, J. Phys. Soc. Jpn. {\bf 63}, 1871 (1994).
%\bibitem{Shih} 
\end{thebibliography}
\end{document}